\documentclass[showpacs,aps,prd,reprint,superscriptaddress,nofootinbib,longbibliography]{revtex4-1}
\usepackage[colorlinks=true, pdfstartview=FitV, linkcolor=red,citecolor=blue, urlcolor=blue,
bookmarks=true, bookmarksnumbered=true, breaklinks]{hyperref}
\usepackage[dvipdfmx]{graphicx}

\usepackage{amsmath,amssymb,tabularx,ulem,mathrsfs,bm,multirow}
\usepackage{braket}

\newcommand*{\e}[1]{
\begin{eqnarray}
#1
\end{eqnarray}
}

\newcommand{\n}[0]{
\nonumber\\}

\begin{document}

\title{Survival probabilities of charmonia as a clue to measure transient magnetic fields}

\author{Sachio Iwasaki}
\email{sutch.iwasaki@th.phys.titech.ac.jp}
\thanks{Corresponding author}
\affiliation{Department of Physics, Tokyo Institute of Technology, Meguro, Tokyo, 152-8551, Japan}
\author{Daisuke Jido}
\email{jido@th.phys.titech.ac.jp}
\affiliation{Department of Physics, Tokyo Institute of Technology, Meguro, Tokyo, 152-8551, Japan}
\author{Makoto Oka}
\email{oka@post.j-parc.jp}
\affiliation{Advanced Science Research Center, Japan Atomic Energy Agency, Tokai, Ibaraki, 319-1195, Japan}
\affiliation{Nishina Center for Accelerator-Based Science, RIKEN, Wako 351-0198, Japan}
\author{Kei Suzuki}
\email{k.suzuki.2010@th.phys.titech.ac.jp}
\affiliation{Advanced Science Research Center, Japan Atomic Energy Agency, Tokai, Ibaraki, 319-1195, Japan}

\date{\today}

\begin{abstract}
We investigate time evolution of $S$-wave charmonium populations under a time-dependent homogeneous magnetic field and evaluate survival probabilities of the low-lying charmonia to the goal of estimating the magnetic field strength at heavy-ion collisions.
Our approach implements mixing between different spin eigenstates and transitions to radially excited states.
We show that the survival probabilities can change even by an extremely short magnetic field.
Furthermore, we find that the survival probabilities depend on the initial spin states.
We propose the sum of the survival probabilities over spin partners as an observable insensitive to the initial states.
We also find that the sum can be approximately given as a function of $\sigma B_0^2$ with a duration time $\sigma$ and the maximum strength of the magnetic field $B_0$.
\end{abstract}
\maketitle

\section{Introduction}
Heavy-ion collision experiments at the Large Hadron Collider (LHC) and the Relativistic Heavy Ion Collider (RHIC) have been operated extensively to uncover the hidden properties of quantum chromodynamics (QCD).
These experiments are supposed to realize various extreme environments, such as high temperature, high-momentum jets, vorticity, and ultra-strong magnetic fields.
The source of the strong magnetic fields is regarded due to the Li\'{e}nard-Wiechert potential from moving charged nuclei.
This simple mechanism would generate the strongest magnetic fields in the current universe, whose amplitude is estimated as $|eB|\sim 50 m_\pi^2\sim 1\ \mathrm{GeV}^2\sim 10^{19}\ \mathrm{Gauss}$ \cite{Rafelski:1975rf,Kharzeev:2007jp,Skokov:2009qp,Voronyuk:2011jd,Ou:2011fm,Bzdak:2011yy,Deng:2012pc,Bloczynski:2012en,Bloczynski:2013mca,Deng:2014uja,Huang:2015oca,Zhao:2017rpf,Zhao:2019crj,Cheng:2019qsn}.
This is much larger than that on the surface of neutron stars, $|eB|\sim 10^{15}\ \mathrm{Gauss}$ \cite{Duncan:1992hi}.
Such a strong magnetic field is a key factor to understand interesting phenomena such as the chiral magnetic effect \cite{Vilenkin:1980fu,Kharzeev:2004ey,Kharzeev:2007tn,Kharzeev:2007jp,Fukushima:2008xe} and the magnetic catalysis of $\braket{\bar qq}$ \cite{Klevansky:1989vi,Suganuma:1990nn,Klimenko:1990rh,Klimenko:1991he,Klimenko:1992ch,Gusynin:1994re,Gusynin:1994va,Gusynin:1994xp,Gusynin:1995nb} which is the enhancement of (the absolute value of) $\braket{\bar qq}$ condensate indicating the spontaneous chiral symmetry breaking in the QCD vacuum.
However, magnetic fields in HICs have never been measured, and it is important to find a probe of magnetic fields.

In experiments or in nature, a magnetic field often depends on time.
For example, the duration time of magnetic fields which are produced in HICs depends on its collision energy and is expected to be extremely short.
A typical duration time of magnetic fields in RHIC is around $\sim 0.1\ \mathrm{fm/c}\sim 10^{-24}\ \mathrm{s}$ \cite{Skokov:2009qp}.
Hence, to measure these extremely transient magnetic fields, we have to focus on particles that are produced within such a short time scale.
Charmonia, bound states of a charm quark and its antiquark, are candidates of such particles because they can be produced in the initial stage of the collision.

Properties of charmonia in a constant magnetic field have been studied by the constituent quark model~\cite{Alford:2013jva,Bonati:2015dka,Suzuki:2016kcs,Yoshida:2016xgm,Iwasaki:2018pby}, QCD sum rules~\cite{Cho:2014exa,Cho:2014loa}, and also effective Lagrangian methods~\cite{Cho:2014exa,Cho:2014loa,Yoshida:2016xgm,Mishra:2020kts}.
There are some characteristic phenomena: (i) the Landau levels of charm quarks (or squeezing of spatial wave function), (ii) the mixing between spin-singlet and spin-triplet eigenstates~\cite{Yang:2011cz,Alford:2013jva,Guo:2015nsa,Bonati:2015dka,Suzuki:2016kcs,Yoshida:2016xgm,Suzuki:2016fof,Dutta:2017pya,Hoelck:2017dby,Iwasaki:2018czv,Iwasaki:2018pby,Chen:2020xsr,Cho:2014exa,Cho:2014loa,Mishra:2020kts} due to the Zeeman coupling of the quarks, (iii) anisotropic (or modified) confinement potential~\cite{Miransky:2002rp,Chernodub:2014uua,Andreichikov:2012xe,Bonati:2014ksa,Rougemont:2014efa,Simonov:2015yka,Bonati:2015dka,Bonati:2016kxj,Bonati:2017uvz,Hasan:2017fmf,Singh:2017nfa,Hasan:2018kvx,Bagchi:2018mdi,Bonati:2018uwh,Khan:2020gky,Hasan:2020iwa,Zhou:2020ssi}, and (iv) the motional Stark effect (or Lorentz ionization) in moving charmonia~\cite{Marasinghe:2011bt,Tuchin:2013ie,Alford:2013jva,Bonati:2015dka,Guo:2015nsa,Chen:2020xsr}.
See Refs.~\cite{Tuchin:2011cg,Machado:2013rta,Dudal:2014jfa,Sadofyev:2015hxa,Braga:2018zlu,Braga:2019yeh,Braga:2020hhs} for other studies and Refs.~\cite{Hattori:2016emy,Zhao:2020jqu,Iwasaki:2021nrz} for reviews.
On the other hand, there are few studies focusing on the time evolution of charmonia in a (time-dependent or constant) magnetic field~\cite{Guo:2015nsa,Suzuki:2016fof,Dutta:2017pya,Hoelck:2017dby,Bagchi:2018mdi}.

In this paper, we investigate the time-evolutions of the low-lying charmonia states in a rapidly varying magnetic field.
We point out the following new phenomena induced by time-dependent magnetic fields:
\begin{enumerate}
\item {\it Time dependence of spin mixing}
---The mixing between different spin eigenstates (e.g., $\eta_c$ and $J/\psi$) by a magnetic field has been well studied.
In this paper we focus on the time evolution of the mixing.
Such an effect was not carefully examined in the previous analyses of time evolution of charmonia in Refs.~\cite{Guo:2015nsa,Hoelck:2017dby,Bagchi:2018mdi}.
\item {\it Suppression of yields of lower states by radial excitation}---Another important effect is a radial excitation from lower states to higher states (e.g., from $\eta_c(1S)$ to $\eta_c(2S)$) which is induced by the quark Landau levels.
This effect finally leads to yield suppression of final states, which will be useful for observing the effects of magnetic fields in experiments.
\item {\it Dependence on initial spin configurations}---Furthermore, we investigate the dependence on the initial spin states, such as the spin-singlet and spin-triplet.
In particular, we find that the survival probabilities are sensitive to initial spin configurations with a complex phase connecting the spin-singlet and spin-triplet states.
Because it may be difficult to determine initial spin configurations experimentally, we propose an independent observable on the initial configuration.
\end{enumerate}
As the final goal, these properties would be helpful to inversely infer the magnetic field strength at HICs.

This paper is organized as follows;
In Sec.~\ref{sec:eqtosolve}, we show the formalism and the numerical setup.
In Sec.~\ref{seq:precession}, we discuss the precession between two levels by comparing numerical results with analytical solutions.
Survival probabilities of the low-lying $S$-wave charmonia are studied numerically in Sec.~\ref{seq:results}.
In Sec.~\ref{sec:scaleParameter}, we derive a scale parameter for the survival probabilities.
Conclusions are given in Sec.~\ref{sec:coclusion}.

\section{Formalism}
\label{sec:eqtosolve}

In this work, we focus on the $S$-wave charmonia, which are the most stable and easiest to experimentally observe among the $c{\bar c}$ states.
They can be handled safely in a non-relativistic framework~\cite{Alford:2013jva,Bonati:2015dka,Suzuki:2016kcs,Yoshida:2016xgm,Iwasaki:2018pby}.
Therefore we consider the time-dependent Schr\"odinger equation.
Since the size of the spatial wave function of the charmonium ground state is smaller than 1 fm, magnetic fields with a wide spatial distribution are well approximated to be homogeneous. 
We take the $z$-axis as the parallel direction to the magnetic field at a time $t$: ${\boldsymbol B}(t)=B(t){\boldsymbol e}_z$.
For simplicity, we assume that the expectation value of the kinetic momentum of the charmonium is zero, namely the pseudomomentum is also zero~\cite{Avron:1978,Alford:2013jva}.
This assumption enables us to separate the center-of-mass coordinate and to express the relative Hamiltonian for the charmonium as follows:
\e{
{H}_{\mathrm{0}} 
&=& 
		-\frac1{m_c} \nabla^2 
	+ b r 
	- \frac 43 \frac{\alpha_s}r
	\n
	&&
	+\frac{32\pi \alpha_s}{9m_c^2}\left( 
		\frac\Lambda{\sqrt{\pi}}\right)^3 
	\left( \mbox{\boldmath $ S$}_1\cdot \mbox{\boldmath $ S$}_2 \right) e^{-\Lambda^2r^2}
+2m_c, \label{eq.diag}
\label{eq.magneticmoment}
\\
V(t)
&=&
\frac{q^2 B(t)^2}{4m_c}\rho^2 
-\sum_{j=1}^2 \left[ \mbox{\boldmath $\mu$}_j \cdot \mbox{\boldmath $B$}(t) \right]
\n
&=&
\frac{q^2 B(t)^2}{4m_c}\rho^2 
- \frac{gq}{2m_c} (\mbox{\boldmath $S$}_1 - {\boldsymbol S}_2)\cdot \mbox{\boldmath $B$}(t)
\label{eq:interaction}
,}
where $m_c$ is the constituent quark mass of the charm quark, $r$ is the distance between the two quarks, and $b$, $\alpha_s$, and $\Lambda$ are model parameters.
${\boldsymbol S}_i$ is the spin operator for the $i$-th quark, where the first one is a charm quark, and the second one is an anti-charm quark.
$\boldsymbol\mu_i=g q_i {\boldsymbol S}_i/(2m_c)$ is the magnetic moment of the $i$-th quark, where the electric charge is $q\equiv q_1=-q_2$, and $g$ is the Land\'e $g$-factor.
We implement the cylindrical coordinate: the $z$-axis is taken as the direction of ${\boldsymbol B}$, and $\rho$ is the perpendicular component to $z$.

The first term of $V(t)$ is derived from the square of the symmetric gauge
(see Ref.~\cite{Alford:2013jva}).
In this work the first term in Eq.~(\ref{eq:interaction}) is called {\it Landau level term}, and the second {\it Zeeman term}.
The Zeeman term induces the mixing between a spin-singlet state and the $S_z=0$ component of a spin-triplet state:
\e{
S_{1z} \left(
	\ket{\uparrow \downarrow}+\ket{ \downarrow\uparrow}
\right)
&=&
+\frac12 \left(
	\ket{\uparrow \downarrow}-\ket{ \downarrow\uparrow}
\right),
\\
S_{2z} \left(
	\ket{\uparrow \downarrow}-\ket{ \downarrow\uparrow}
\right)
&=&
-\frac12 \left(
	\ket{\uparrow \downarrow}+\ket{ \downarrow\uparrow}
\right)
.}
Note that the $S_z=\pm1$ components of the spin-triplet state are not mixed with other states via the Zeeman term, so that we do not consider them in this work.

We expand the time-dependent wave function with the eigenfunctions of the in-vacuum Hamiltonian $H_0$ as follows:
\e{
\Psi = \Psi({\boldsymbol r}; t) = \sum_{j=1}^N e^{-i E_j t} c_{j}(t) \Psi_j({\boldsymbol r}),
}
where $E_j$ and $\Psi_j$ are the $j$-th eigenvalue and eigenfunction of $H_0$, respectively: $H_0 \Psi_j = E_j \Psi_j$.
$N$ is the number of basis functions to expand a time-dependent state.
We use only the $S$-wave states because the deformation of spatial wave functions under a finite magnetic field, namely mixture of higher orbital angular momenta such as $L=2,4,\cdots$, is expected to be small.
$c_j(t)$ is an expansion coefficient, and we call $|c_j(t)|^2$ {\it population} of the $j$-th state.
Using the normalization condition, $\braket{\Psi_i|\Psi_j} = \delta_{ij}$, we obtain a set of equations to solve for the populations:
\e{
\dot c_{j}(t)
=
-i \sum_k^N V_{jk} e^{i(E_j - E_k)t} c_{k}(t),
\label{eq:eqtosolve}
}
where 
$V_{jk} := \Braket{\Psi_j| V(t) | \Psi_k}$.
We solve Eq.~(\ref{eq:eqtosolve}) with the wave functions and their corresponding eigenenergies of the $S$-wave charmonia that are obtained from Refs.~\cite{Suzuki:2016kcs,Yoshida:2016xgm} using the cylindrical Gaussian expansion method based on the constituent quark model.
We set $m_c=1.784\ \mathrm{GeV}$, $\alpha_s = 0.713$, and the string tension $\sqrt{b}= 0.402$ GeV from Ref.~\cite{Kawanai:2015tga} obtained by the equal-time $Q{\bar Q}$ Bethe-Salpeter amplitude simulated in the lattice QCD.
We use $\Lambda=$1.020 GeV$^2$ as a Gaussian-type smearing parameter for the spin-spin potential of $H_0$, where its value is obtained by fitting the quark-antiquark potential measured in lattice QCD simulations in Ref.~\cite{Kawanai:2011jt}.

In Ref.~\cite{Deng:2012pc}, Deng and Huang estimated strengths of magnetic fields by using numerical simulations with the Heavy Ion Jet Interaction Generator (HIJING) model which is a Monte-Carlo event generator for high-energy collisions.
In peripheral collisions at LHC, magnetic-field strengths can reach $70m_\pi^2 \gtrsim 1.3 \ \mathrm{GeV^2}$.
In this work, we cover the region of the order of 1 GeV$^2$.

\section{two-level precession under constant magnetic field}
\label{seq:precession}

\begin{figure}[t!]
    \centering
    \includegraphics[width=1.0\columnwidth]{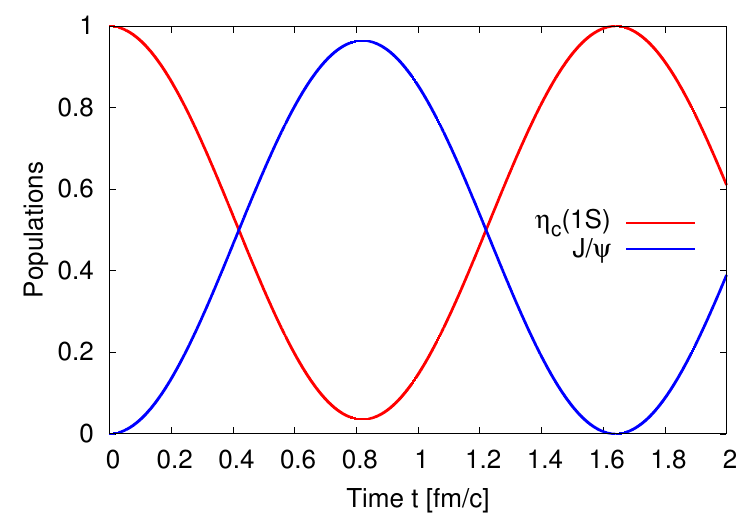}
    \caption{Time evolution of populations of the 1S charmonia, $\eta_c(1S)$ and $J/\psi (S_z=0)$, under a constant external magnetic field with $|eB|=1\ \mathrm{GeV^2}$.
The initial state is occupied by only $\eta_c(1S)$.
The number of basis functions is $N=2$, which is a pedagogical two-level system.}
    \label{Fig:2levelsConst}
\end{figure}

In this section, we discuss a precession period of populations for a simple two-level system ($N=2$) under a constant magnetic field.
We consider the system composed of the two lowest charmonium states of $\eta_c(1S)$-$J/\psi(S_z=0)$.
In this case, populations vary as sine waves.
In Fig.~\ref{Fig:2levelsConst}, we show the numerical solutions of time-dependent populations from Eq.~(\ref{eq:eqtosolve}) under $|eB|=1\ \mathrm{GeV^2}$ as an example.
The matrix elements $V_{jk}$ and eigenenergies $E_j$ are taken from a constituent quark model calculation in Refs.~\cite{Yoshida:2016xgm,Suzuki:2016kcs}, where the energy difference between two levels is $\Delta E= E_2-E_1=$ 113.3 MeV.
From the spin states of the basis, we obtain off-diagonal matrix elements as below:
\e{
|V_{12}|
=
\Braket{\eta_c(1S)|
\frac{gq}{2m_c} (\mbox{\boldmath $S$}_1 - {\boldsymbol S}_2)\cdot \mbox{\boldmath $B$}
|J/\psi}
\sim
\frac{gqB}{2m_c}
\label{eq:matrixElement}
,\hspace{0.5cm} }where the approximation in the last line
comes from the assumption that the spin partners have similar spatial wave functions.
For the parameters of the matrix elements, we take $g=2$, and $q=(2/3)e$.

We can derive the analytical solution for the precession period $T$ by a straightforward calculation from Eq.~(\ref{eq:eqtosolve}) with $N=2$:
\e{
T=
\frac{2\pi}{\sqrt{(V_{22}-V_{11}+\Delta E)^2+4 |V_{12}|^2}}
\label{eq:period}
.}
The two-level system has diagonal matrix elements $V_{11}$ and $V_{22}$, but their effects cancel as seen in Eq.~(\ref{eq:period}) with the form of $V_{22}-V_{11}$.
Inputting the parameters above, we can estimate the precession period as follows:
\e{
T (|eB|=1 \ \mathrm{GeV}^2)
=
\frac{2\pi}{\sqrt{(\Delta E)^2+4 |V_{12}|^2}}
\sim
1.64\ \mathrm{fm/c}.\nonumber\\
}
This period reproduces the numerical one seen in Fig.~\ref{Fig:2levelsConst}.

\begin{figure}[t!]
    \centering
    \includegraphics[width=1.0\columnwidth]{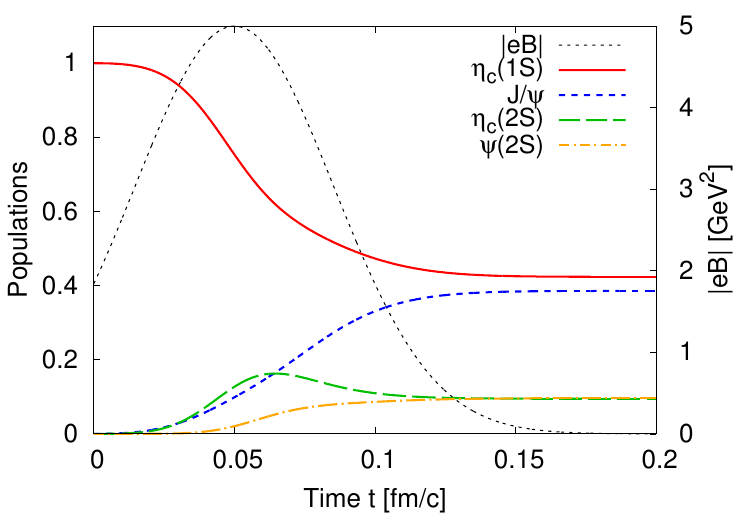}
     \caption{Time evolution of populations for the $S$-wave charmonia, $\eta_c(1S)$ and $J/\psi (S_z=0)$, under a time-dependent external magnetic field with $|eB_0|=5\ \mathrm{GeV^2}, \sigma=0.05\ \mathrm{fm/c},$ and $t_{\mathrm{peak}}=0.05$ fm/c.
The initial state is occupied by only $\eta_c(1S)$.
The number of basis functions is $N=4$, which is a pedagogical four-level system.}
    \label{Fig:2levelsGauss}
\end{figure}

\section{multi-level precession under time-dependent magnetic field}
\label{seq:results}
From this section we consider time-dependent magnetic fields.
As in the case of HICs, a rapid change of the magnetic field will mix the ground states with higher excited states. As a consequence, the population of the ground states, $\eta_c(1S)$ or $J/\psi$, will be reduced by the magnetic field.
As a typical time dependence of a magnetic field produced in HICs, we here consider a Gaussian form:
\e{
B(t)
=
B_0
\exp{
	\left[
		-\frac{(t-t_{\mathrm{peak}})^2}{\sigma^2}
	\right]
	}
.}
This form is characterized by the three parameters, $B_0$, $\sigma$, and $t_{\mathrm{peak}}$.
$t_{\mathrm{peak}}$ is an ``offset time" which reflects that the strength of the magnetic field is maximum at a certain time later than the first contact of two heavy ions with a finite size.

Magnetic fields cause two types of mixing:
(i) between a spin-singlet state and the $S_z=0$ components of a spin-triplet state, and (ii) between radial excitation states.
Figure \ref{Fig:2levelsGauss} shows time evolution of populations for the 1S and 2S charmonium states within a four-level system ($N=4$) as a pedagogical example.
As an initial state, we assume that the $\eta_c(1S)$ state occupies 100 \% of the population.
One sees that the time-dependent magnetic field induces not only the mixing of the spin partner $J/\psi$ state, but also the mixing of the radially 2S excited states, $\eta_c(2S)$ and $\psi(2S)$.
The mixings of the higher excited states grow as the maximum strength of the magnetic field increases.

\begin{figure}[t!]
    \centering
    \includegraphics[width=1.0\columnwidth]{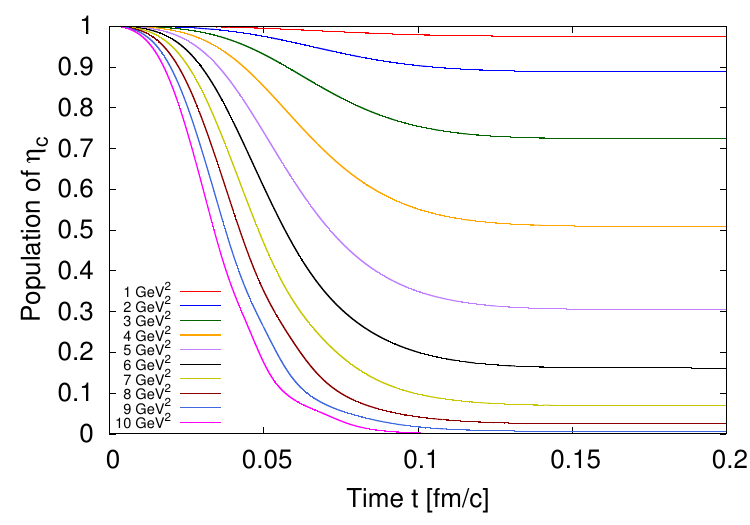}
    \caption{Time evolution of populations of the $\eta_c (1S)$ undergoing a Gaussian magnetic field with the maximum strength of 1-10 GeV$^2$.
Other parameters are fixed as $\sigma=0.05$ fm/c, and $t_{\mathrm{peak}}=0.05$ fm/c.
The initial state is occupied by only $\eta_c(1S)$.
The number of basis functions is $N=20$.
    }
    \label{Fig:1Spopulation}
\end{figure}

Figure~\ref{Fig:1Spopulation} shows time evolution of the $\eta_c(1S)$ populations under different maxima of magnetic fields, where $B_0$ runs from 1 GeV$^2$ to 10 GeV$^2$.
In this calculation, the number of basis functions is $N=20$, which is sufficient for the convergence of numerical results of the low-lying states.
We can see that the $\eta_c(1S)$ populations are gradually suppressed with larger $B_0$.
After the magnetic field vanishes, the population becomes constant.
We call this remaining value {\it survival probability}.
This is a physical quantity characterized by a transient magnetic field as in experiments.

\begin{figure}[t!]
    \centering
    \includegraphics[width=1.0\columnwidth]{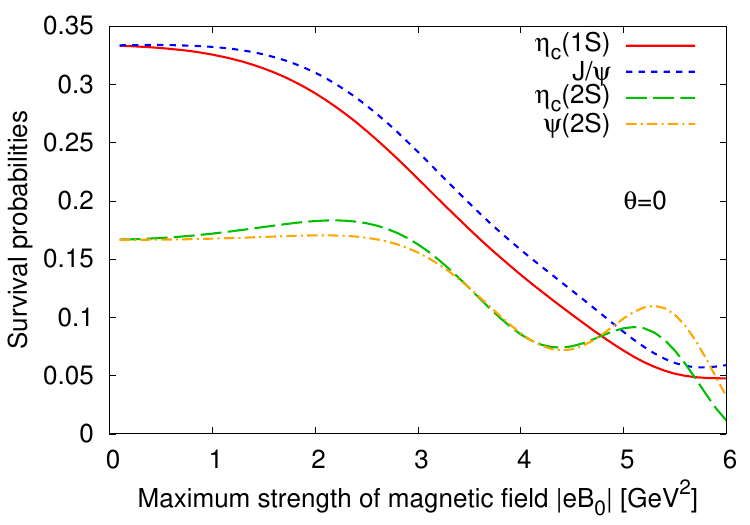}
    \caption{
Survival probabilities of charmonia, where the horizontal axis is $|eB_0|$.
Other parameters are fixed as $\sigma=0.05$ fm/c, and $t_{\mathrm{peak}}=0.05$ fm/c.
The initial state starts from $|c_{\eta_c(1S)}|^2=|c_{J/\psi}|^2=2|c_{\eta_c(2S)}|^2=2|c_{\psi(2S)}|^2=1/3$, where $\theta=0$ is fixed.
The number of basis functions is $N=20$.}
    \label{fig:surPro_s0.05_mixed}
\end{figure}

Figure~\ref{fig:surPro_s0.05_mixed} shows survival probabilities of the 1S and 2S charmonia against different $B_0$.
As the initial condition, we implement that
$|c_{\eta_c}|^2: |c_{J/\Psi}|^2: |c_{\eta_c(\mathrm{2S})}|^2: |c_{\psi(2S)}|^2=1:1:0.5:0.5$, which corresponds to the typical magnitudes of the wave functions at the origin.
Here we fix only the absolute values of the population, and 
we treat the complex phase difference among $c_j$ as a parameter.
We express initial states as 
$\ket{\mathrm{spin\ singlet}} + e^{i\theta} \ket{\mathrm{spin\ triplet}}$, namely $\ket{\uparrow \downarrow- \downarrow \uparrow} + e^{i\theta} \ket{\uparrow \downarrow + \downarrow \uparrow}$, where the normalization factor is omitted.
Here $\theta$ determines the initial spin state: e.g., $\theta=0$ corresponds to $\ket{\uparrow \downarrow}$, while $\theta=\pi$ is $\ket{\downarrow\uparrow}$.

\begin{figure}[t!]
    \centering
    \includegraphics[width=1.0\columnwidth]{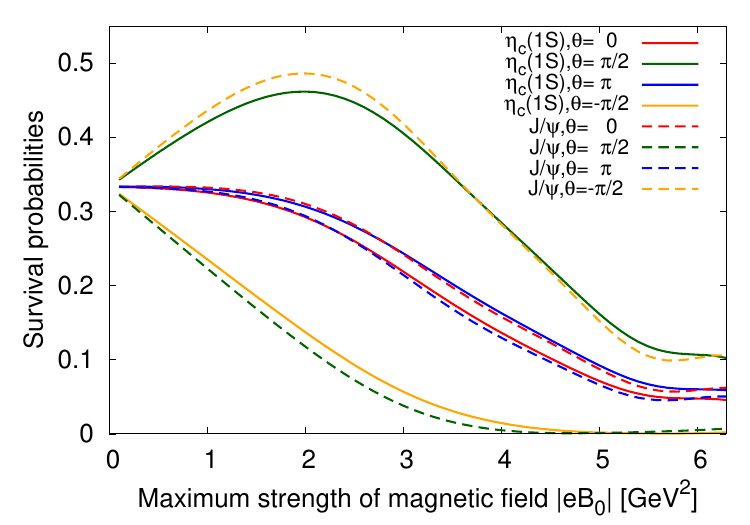}
    \caption{Survival probabilities of the 1S charmonia, $\eta_c(1S)$ and $J/\psi (S_z=0)$, where the horizontal axis is $|eB_0|$.
    Other parameters except for $\theta$ are the same as Fig.~\ref{fig:surPro_s0.05_mixed}.}
    \label{fig:diffTheta}
\end{figure}

\begin{figure}[t!]
    \centering
    \includegraphics[width=1.0\columnwidth]{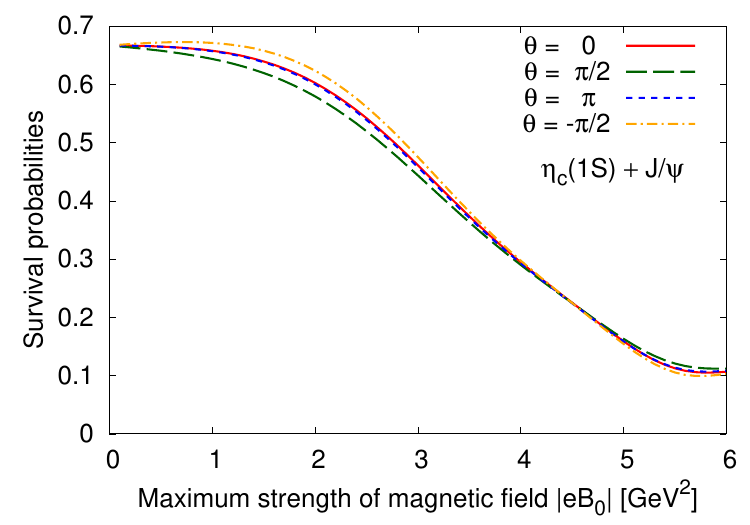}
    \caption{
Sum of the survival probabilities of the 1S charmonia, $\eta_c(1S)$ and $J/\psi (S_z=0)$, where the horizontal axis is $|eB_0|$.
Other parameters except for $\theta$ are the same as Fig.~\ref{fig:surPro_s0.05_mixed}.
}
    \label{fig:6GeV}
\end{figure}

In Fig.~\ref{fig:diffTheta}, we show the survival probabilities for various choices of the initial phase $\theta$.
From this figure, we find that the survival probabilities significantly depend on $\theta$, or initial spin states, which is mostly caused by the Zeeman effect for the quark spins.
If one could determine the initial $\theta$ in experiments, the $\theta$ dependence would be important as a probe of the magnetic field.
In reality, it may be more difficult to experimentally infer the initial $\theta$ than the strength and duration time of the magnetic field.
Then it is useful to provide an observable independent of the initial $\theta$.
Here, we propose that the sum of the survival probabilities over spin partners is a useful quantity.
Figure~\ref{fig:6GeV} shows the sum of survival probabilities for the 1S charmonia, $\eta_c(1S)$ and $J/\psi$, for the same setup as Fig.~\ref{fig:diffTheta}.
With each $B_0$, the sum is almost invariant with respect to the different $\theta$, namely different initial spin configuration.
In this setup, the maximum of the relative error between them is around 2 \%.
Thus, we can conclude that {\it the spin sum of the survival probabilities will be a useful observable independent of initial $\theta$ in experiments}.

\section{Scale parameter of survival probabilities}
\label{sec:scaleParameter}
\begin{figure}[t!]
    \centering
    \includegraphics[width=1.0\columnwidth]{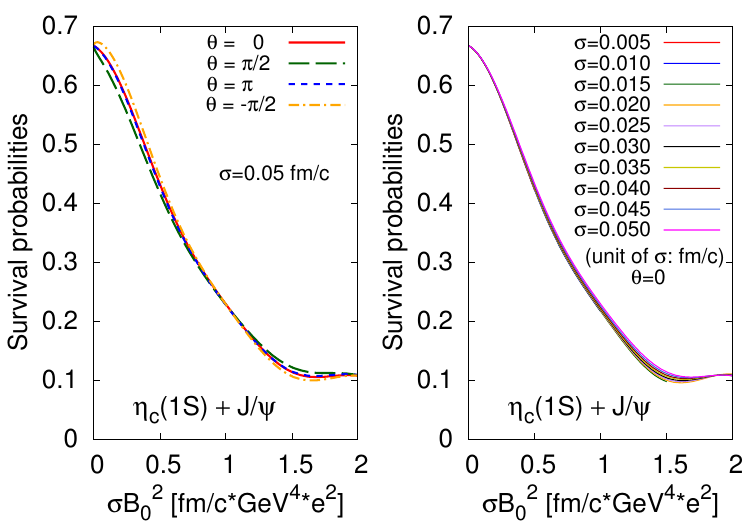}
    \caption{
Sum of survival probabilities of the 1S charmonia, $\eta_c(1S)$ and $J/\psi (S_z=0)$, {\it where the horizontal axis is the scale parameter $\sigma B_0^2$}.
Other parameters except for $\sigma$ and $\theta$ are the same as Fig.~\ref{fig:surPro_s0.05_mixed}.
Left: Comparison with different $\theta$ at $\sigma=0.05$ fm/c.
Right: Comparison with different $\sigma$ at $\theta=0$.
	}
    \label{fig:s1Ssummed}
\end{figure}

In Fig.~\ref{fig:6GeV}, we have shown the $B_0$ dependence of the survival probabilities which is almost independent of $\theta$.
In this section, we propose a new parameter $\sigma B_0^2$, which we call {\it scale parameter}.
In Fig.~\ref{fig:s1Ssummed}, we show the $\sigma B_0^2$ dependence of the same quantities.
The left and right panels correspond to different $\theta$ at a fixed $\sigma$ and different $\sigma$ at a fixed $\theta$, respectively.
From these results, we can conclude that {\it the spin sum of the survival probabilities is scaled by $\sigma B_0^2$, and this scaling is almost independent of both $\theta$ and $\sigma$}.
Therefore, even though the experimental determination of $\theta$ and $\sigma$ may be difficult, we can know $\sigma B_0^2$ from the observation of the spin sum of the survival probabilities.

In what follows, we derive that $\sigma B_0^2$ dominates the suppression of the spin sum of the survival probabilities by a first-order perturbation.
For simplicity, we consider a two-level system consisting of a state and its radially excited state.
It is straightforward to include more levels.
As a first-order perturbation, we approximate $c(t)\sim c(0)$ in Eq.~(\ref{eq:eqtosolve}):
\e{
\dot c_{1}(t)
=
V_{11}(t) c_1
-i V_{12}(t) e^{i(E_1 - E_2)t} c_{2},
\label{eq:c1eq}
}
where the index $1$ or $2$ means a radial quantum number.
Although we have written only the equation for the first element $\dot c_{1}(t)$, that for the second element $\dot c_{2}(t)$ also has a similar form.

Integrating Eq.~(\ref{eq:c1eq}) from $t=0$ to $t=\infty$, we obtain
\e{
c_1(\infty)
=
c_1(0)
+\int_0^\infty \mathrm{d}t 
\left(
	V_{11}(t) c_1
	-i
	V_{12}(t) e^{i\Delta Et} c_2
\right). \label{eq:c1inf}
\hspace{5mm}}
Here we focus on only the calculation for the term with $V_{11}(t)$.
For $V_{11}(t)$, we use the Landau level term in Eq.~(\ref{eq:interaction}).
Then,
\e{
&&
\int_0^\infty dt V_{11}(t) c_1
\n 
&\sim&
	\int_{-\infty}^\infty \mathrm dt 
	\Braket{\Psi_1|
		\frac{q^2 B_0^2 \rho^2 }{8\mu} \exp\left(-\frac{2(t-t_0)^2}{\sigma^2} \right)
		|\Psi_1}
		c_1
\n 
&=&
\frac{q^2}{8\mu}
\sqrt{\frac{\pi}2}
\Braket{\Psi_1| \rho^2 |\Psi_1}
\sigma B_0^2 c_1
,
}
where we have approximated the integration interval to $(-\infty,\infty)$ in the second line by assuming that the Gaussian tail before the initial time $t=0$ is small.
As a result, we can obtain $\sigma B_0^2$ as a constant of proportionality for the suppression of survival probabilities.
Note that for the term with $V_{12}(t)$ in Eq.~(\ref{eq:c1inf}), we consider only the Landau level term in Eq.~(\ref{eq:interaction}) and neglect the Zeeman term.
This is because the latter gives little overlap between different radial states, such as $\eta_c(1S)$-$\psi(2S)$.
Then we obtain an additional exponential factor of $\exp[ -(\sigma\Delta E)^2/8]$ as a constant of proportionality.
However, by considering typical values of $\sigma$ and $\Delta E$, we can ignore this factor.
For example, the typical scale of $\sigma$ expected in experiments is approximately 0.01 fm/c.
$\Delta E$ is approximately 500 MeV since we consider a radial excitation of charmonia.
From these values, we can estimate $\sigma^2(\Delta E)^2\sim 1/1600$ and $\exp[ -\sigma^2(\Delta E)^2/8] \sim 0.99992$.
Hence this exponential factor can be ignored as approximately $1$.
Thus, we can conclude that $\sigma B_0^2$ behaves as a scale parameter for both the $V_{11}(t)$ and $V_{12}(t)$ terms.

\section{Conclusion}
\label{sec:coclusion}

The time-dependent behaviors of the $S$-wave charmonia under a rapidly changing magnetic field have been studied.
We have calculated the spin mixings and the radial excitations of the charmonia during the exposure to the magnetic field.
We have shown the survival probabilities of the low-lying charmonium states as functions of the parameters of time-dependent magnetic fields.
As a result, we have the following conclusions:
\begin{enumerate}
\item We have clarified time evolution of charmonium populations by the spin mixings and the radial excitations under time-dependent magnetic fields.
\item We have shown that yields of lower states are suppressed.
Low-lying states which are produced as initial states make transitions to radially-excited higher states, and their yields are suppressed.
\item We have found that the survival probability of {\it one state} depends on initial spin configurations characterized by the parameter $\theta$.
It might be difficult to experimentally detect the initial $\theta$, but the sum of survival probabilities of the spin-singlet and spin-triplet states is not sensitive to the choice of the initial $\theta$.
Thus this may be a good candidate for measuring the magnetic field.
\item We have found that the spin sum of the survival probabilities scales with a product of the duration time and the squared maximum strength of the magnetic field, $\sigma B_0^2$.
\end{enumerate}
Even if the duration time of magnetic fields is extremely transient, the survival probabilities of the charmonia would change significantly.
This effect would help to estimate the magnetic field in HICs, and at least it give a possibility to infer $\sigma B_0^2$ from the suppression for the sum of the survival probabilities over spin partners.

Finite-temperature effects in HICs can influence various observables related to charmonia.
In particular, the survival probabilities are expected to be suppressed by the dissociation of charmonia inside the quark gluon plasma (QGP), which is the so-called ``charmonium suppression."
However, collisions creating strong magnetic fields, as considered in this work, are {\it ultraperipheral}.
Temperature after such collisions may be lower than that of the QGP, so that resultant charmonium suppression is expected to be insignificant.
If we consider charmonia at higher temperature, we have to take into account thermal deformation of the potential, such as the Debye screening,  and dissociation process into quark-antiquark pairs.

Furthermore, in general, charmonia produced in HICs can move with a nonzero momentum.
The center-of-mass momentum in a magnetic field is not conserved, but the pseudomomentum $\mbox{\boldmath $K$}$ is instead conserved.
When $\mbox{\boldmath $K$}$ is nonzero, we have to add the term proportional to $(\mbox{\boldmath $K$} \times \mbox{\boldmath $B$}(t)) \cdot \mbox{\boldmath $r$} $ to Eq.~(\ref{eq:interaction}) \cite{Alford:2013jva}.
From this form, when we focus only on $\mbox{\boldmath $K$}$ parallel to $\mbox{\boldmath $B$}$ (namely, the $z$-direction), this term is zero.
In other words, charmonia moving in the $z$-direction are not affected by the pseudomomentum.
In HICs, the direction of magnetic field is mostly perpendicular to the reaction plane of collisions.
Therefore, when charmonia moving in this direction are selectively measured, our analysis will be relevant.
On the other hand, if $\mbox{\boldmath $K$}$ is not parallel to $\mbox{\boldmath $B$}$, the pseudomomentum can influence the survival probabilities, and such a calculation would be interesting for future studies.

In our calculation, we have used only the $S$-waves at zero magnetic field as basis functions and have truncated the states with a nonzero orbital angular momentum ($L=2,4, \cdots$).
It would be interesting to include the higher partial waves (or equivalently deformed wave functions as in Refs.~\cite{Suzuki:2016kcs,Yoshida:2016xgm}).

Though this work is motivated by the extreme environment in HIC experiments, our calculation method itself may be useful in not only hadron systems but also arbitrary nonrelativistic quantum systems such as atoms and molecules with time-dependent magnetic fields.
For instance, see Ref.~\cite{PhysRevLett.124.173202} for transition of a positronium in a spatially periodic magnetic field.
Since the charmonium is analogous to the positronium, we can apply our method and discussion directly to them.

\section*{Acknowledgments}
This work is partially supported by the Grant-in-Aid for Scientific Research (Grants Nos.~JP17K05449,
JP17K14277,
JP19H05159,
JP19J13655,
JP20K03959,
and
JP20K14476) from the Japan Society for the Promotion of Science.
S. I. was partially supported by the JAEA student scholarship.

\bibliography{timeDep}
\end{document}